\documentclass[%
superscriptaddress,
nofootinbib,
nobibnotes,
amsmath,amssymb,
aps,
floatfix,
]{revtex4-2}

\usepackage{placeins}
\usepackage{booktabs}
\usepackage{dcolumn}
\usepackage{array}
\usepackage{longtable}
\usepackage{float}
\usepackage{hyphenat}
\usepackage{natbib}
\usepackage{hyperref}

\usepackage{amsmath,amsfonts,amsthm,amssymb}
\usepackage{bm,graphicx,graphics,color}
\usepackage{epsf,epsfig}
\usepackage[all]{xy}
\newcommand*{\B}[1]{\ifmmode\bm{#1}\else\textbf{#1}\fi}

\def\bx{\mathbf{x}}

\def\bk{\mathbf{k}}

\def\no{\nonumber}
\def\lb{\label}
\def\be{\begin{equation}}
\def\ee#1{\label{#1}\end{equation}}
\newcommand{\ben}{\begin{eqnarray}}
\newcommand{\een}{\end{eqnarray}}

\begin{document}

\title{Plane wave analysis of the  second post-Newtonian hydrodynamic equations}

\author{Gilberto M. Kremer}
\email{kremer@fisica.ufpr.br}
\affiliation{Departamento de F\'{i}sica, Universidade Federal do Paran\'{a}, Curitiba 81531-980, Brazil}

%%%%%%%%%%%%%%%%%%%%%%%%%%%%%%%%%%%%%%%%%%%%%%%%%%%%%%%%%%%%%%%
\begin{abstract}
The second post-Newtonian hydrodynamic equations are  analyzed within the framework of a plane wave solution.  The hydrodynamic equations for the mass and momentum density are coupled with six Poisson equations for the Newtonian and post-Newtonian gravitational potentials. Perturbations of the basic fields and gravitational potentials from a background state by assuming plane wave representations lead to a dispersion relation where the Jeans instability condition emerges. The influence of the first and second post-Newtonian approximations on the Jeans mass is determined. It was shown that the relative difference  of the first  post-Newtonian and the Newtonian  Jeans masses is negative while the one of the second post-Newtonian approximation is positive. The two contributions  imply a smaller mass  needed for an overdensity to initiate  the gravitational collapse than the one given by the Newtonian theory.
\end{abstract}

\maketitle
\section{Introduction}
\lb{s.1}

The  determination of self-gravitating fluid instabilities from the hydrodynamic equations coupled with the Newtonian Poisson equation is an old subject in the literature which goes back to the pioneer work of Jeans in 1922  \cite{Jeans}.
The analysis of the instabilities is based on a  dispersion relation where one can infer a wavelength cutoff, known as the Jeans wavelength. Two distinct behaviors follow from the dispersion relation,  one  refers to wavelength perturbations that are smaller than the Jeans wavelength  and the perturbations propagate as harmonic waves in time, while in the other the wavelength perturbations are greater than the Jeans wavelength and  the perturbations grow or decay in time.  The Jeans instability \cite{Wein,Coles,BT1} refers to the  gravitational collapse of  self-gravitating interstellar gas clouds which are  associated with the mass density perturbations which grow exponentially with time.  A simple physical model  is that the collapse of a mass density inhomogeneity happens  if the inwards gravitational force is bigger than the outwards internal pressure force of the self-gravitating interstellar gas cloud.

  The first  post-Newtonian hydrodynamic equations for Eulerian fluids from Einstein's field equations where all terms in the $\mathcal{O}(c^{-2})$ order are considered were derived by Chandrasekhar \cite{Ch1}. Apart from the Newtonian gravitational potential there exist a scalar and a vector gravitational potentials which are coupled with the hydrodynamic equations through Poisson equations.  The second post-Newtonian hydrodynamic equations for Eulerian fluids were derived by Chandrasekhar and Nutku \cite{ChNu} where all terms in the $\mathcal{O}(c^{-4})$  order are taken into account. In this approximation the hydrodynamic equations are coupled with six Poisson equations, since there appear -- additionally to the gravitational potentials of the first post-Newtonian approximation -- a scalar, a vector and a tensor gravitational potentials.

 The Jeans instability  was recently analysed within the framework of the first post-Newtonian theory by two different methodologies. In \cite{NKR,NH} the  hydrodynamic equations for the mass and momentum densities are coupled with the Poisson equations, while in \cite{gg1} the first post-Newtonian expression for the collisionless Boltzmann equation \cite{Ped,gg2,GGKK} is coupled with the Poisson equations. It was shown that in the first post-Newtonian analysis  the mass necessary for an overdensity to begin the gravitational collapse  is smaller than the one in the Newtonian theory.
 
 The Jeans instability was also analysed within the framework of $f(R)$ gravity or modified gravity theories \cite{SC1,SC2,SC3} where a modified dispersion relation along with a new kind of an unstable mode follow.
 
 The aim of this work is to analyse the equations that follow from the second post-Newtonian approximation of Einstein's field equations. In the second post-Newtonian theory   the  hydrodynamic equations for the mass and momentum densities are coupled with six Poisson equations for the Newtonian and post-Newtonian gravitational potentials. Here we analyse the perturbations from a background state of the basic fields and gravitational potentials by assuming  plane wave representations which lead to a system of algebraic equations whose solution is a dispersion relation.  From the dispersion relation   the condition related with the Jean instability comes out and as a consequence it is possible to
 determine the influence of the second post-Newtonian approximation in the Jeans mass, which is related with the minimum mass necessary for an overdensity to initiate the gravitational collapse.

The paper is structured as follows. In Section \ref{s.2} the hydrodynamic equations for the mass and momentum densities and  the six Poisson equations which follow from the  second post-Newtonian approximation are introduced.  The subject of Section \ref{s.3} is to determine the hydrodynamic and Poisson equations when  perturbations in the background state of the  fields and and gravitational potentials are considered.  In Section \ref{s.4}  a dispersion relation is obtained from a plane wave representation of the perturbations where the post-Newtonian influence in the Jeans mass  is analyzed. The conclusions of the work are stated in the last section.

\section{Hydrodynamic and Poisson equations}
\lb{s.2}

The components of the metric tensor  in the second post-Newtonian approximation were derived by Chandrasekhar and Nutku \cite{ChNu} from Einstein's field equations and read  
\ben\lb{sp1a}
&&g_{00}=1-\frac{2U}{c^2}+\frac2{c^4}\left(U^2-2\Phi\right)+ \frac{\Psi_{00}}{c^6}+\mathcal{O}(c^{-8}),
\\\lb{sp1b}
&&g_{0i}=\frac{\Pi_i}{c^3}+\frac{\Psi_{0i}}{c^5}+\mathcal{O}(c^{-7}),
\\\lb{sp1c}
&&g_{ij}=-\left(1+\frac{2U}{c^2}\right)\delta_{ij}+\frac{\Psi_{ij}}{c^4}+\mathcal{O}(c^{-6}),
\een
where the Newtonian   $U$ and the post-Newtonian $\Phi$, $\Pi_i$, $\Psi_{ij},$ $\Psi_{0i}$ and $\Psi_{00}$ gravitational potentials  satisfy the Poisson equations 
\ben\lb{sp2a}
&&\nabla^2U=-4\pi G\rho,\qquad\nabla^2\Phi=-4\pi G\rho\left(V^2+U+\frac\varepsilon2+\frac{3p}{2\rho}\right),
\\\lb{sp2aa}
&&\nabla^2\Pi_i=-16\pi G\rho V_i+\frac{\partial^2U}{\partial t\partial x^i},
\\\lb{sp2b}
&&\nabla^2\Psi_{kk}=32\pi G\rho\left(V^2+4U+\varepsilon\right)-12\bigg(\frac{\partial U}{\partial x^j}\bigg)^2,
\\\lb{sp2c}
&&\nabla^2\Psi_{0i}=-16\pi G\rho\left[V_i\left(V^2+\varepsilon+\frac{p}\rho+4U\right)-\frac{\Pi_i}2\right]-10\frac{\partial U}{\partial t}\frac{\partial U}{\partial x^i}
-2\frac{\partial U}{\partial x^j}\frac{\partial \Pi_j}{\partial x^i}+2\Pi_j\frac{\partial^2 U}{\partial x^i\partial x^j},
\\\no
&&\nabla^2\Psi_{00}=16\pi G\rho\left[V^2\left(V^2+\varepsilon+\frac{p}\rho+4U\right)-U^2-2\Phi\right]
+2\frac{\partial U}{\partial x^i}\frac{\partial \Pi_i}{\partial t}-6\left(\frac{\partial U}{\partial t}\right)^2
\\\lb{sp2d}
&&\qquad+12\frac{\partial U}{\partial x^i}\frac{\partial \Phi}{\partial x^i}+\frac{\partial \Pi_j}{\partial x^i}\left(\frac{\partial \Pi_i}{\partial x^j}-\frac{\partial \Pi_j}{\partial x^i}\right)
-12U\left(\frac{\partial U}{\partial x^i}\right)^2+2\Psi_{ij}\frac{\partial^2 U}{\partial x^i\partial x^j}.
\een
In the above  Poisson equations $G$ is the universal gravitational constant and their expressions  were written by considering the gauge \cite{ChNu}
\ben\lb{sp2e}
3\frac{\partial U}{\partial t}+\frac{\partial \Pi_i}{\partial x^i}+\frac1{c^2}\left[\frac{\partial\Psi_{0j}}{\partial x^j}-\frac12\frac{\partial\Psi_{jj}}{\partial t}\right]=0.
\een
Here we follow \cite{gg2,GGKK} and denote the potentials  of \cite{ChNu} as : $\Pi_i\equiv P_i$, $\Psi_{ij}\equiv Q_{ij}$, $\Psi_{0i}\equiv Q_{0i}$ and $\Psi_{00}\equiv Q_{00}$. Furthermore, in (\ref{sp2a}) -- (\ref{sp2d}) $\rho$ denotes the mass density, $p$ the hydrostatic pressure, $V_i$ the hydrodynamic velocity and $\varepsilon$ the specific internal energy of the fluid.

 In the second post-Newtonian approximation the continuity equation  becomes \cite{ChNu,gg2,GGKK}
\ben\lb{ba2k}
 \frac{\partial\widetilde\rho}{\partial t}+\frac{\partial \widetilde \rho V_i}{\partial x^i}=0,
 \een
 where the mass density $\widetilde\rho$ is given by
\be
 \widetilde\rho=\rho\bigg[1+\frac1{c^2}\bigg(\frac{V^2}2+3U\bigg)+\frac1{c^4}\bigg(\frac38V^4+\frac72UV^2+\frac32U^2-\frac12\Psi_{kk}
 -\Pi_iV_i\bigg)\bigg].
 \ee{ba2ka}

The hydrodynamic equation for the momentum density in the second post-Newtonian approximation  can be written as \cite{ChNu,gg2,GGKK}
\ben\no
&&\frac{\partial \rho{\mathfrak{V}}_i}{\partial t}+\frac{\partial \rho{\mathfrak{V}}_iV_j}{\partial x^j}+\frac{\partial p}{\partial x^i}\left[1+\frac{2U}{c^2}-\frac1{c^4}\left(U^2+2\Phi+\frac{\Psi_{kk}}2\right)\right]-\rho\frac{\partial U}{\partial x^i}\bigg\{1+\frac2{c^2}\bigg(V^2+U+\frac\varepsilon2+\frac{p}{2\rho}\bigg)
\\\no
&&\qquad+\frac2{c^4}\bigg[V^4+5UV^2-\frac{3U^2}2+\Phi-\Pi_iV_i+(V^2+U)\bigg(\varepsilon+\frac{p}\rho\bigg)-\frac{\Psi_{kk}}4\bigg]\bigg\}
\\\lb{ba4h}
&&\qquad
+\frac\rho{c^2}\bigg(V_j\frac{\partial \Pi_j}{\partial x^i}-2\frac{\partial \Phi}{\partial x^i}\bigg)\bigg[1+\frac1{c^2}\bigg(V^2
+4U+\varepsilon+\frac{p}\rho\bigg)\bigg]+\frac\rho{2c^4}\bigg(\frac{\partial \Psi_{00}}{\partial x^i}+2V_j\frac{\partial \Psi_{0j}}{\partial x^i}+V_jV_k\frac{\partial \Psi_{jk}}{\partial x^i}\bigg)=0,
\een
where the following abbreviation for the momentum density was introduced
\ben\no
&&\rho{\mathfrak{V}}_i=\rho V_i\bigg\{1+\frac1{c^2}\bigg(V^2+6U+\varepsilon+\frac{p}\rho\bigg)+\frac1{c^4}\bigg[V^4+10V^2U+2\Phi-2\Pi_iV_i-\frac{\Psi_{kk}}2
+13U^2
\\\lb{ba4i}
&&\quad+\big(V^2+6U\big)\bigg(\varepsilon+\frac{p}\rho\bigg)\bigg]\bigg\}-\frac\rho{c^2}\Pi_i\bigg[1+\frac1{c^2}\bigg(V^2
+4U+\varepsilon+\frac{p}\rho\bigg)\bigg]-\frac\rho{c^4}\big(\Psi_{0i}+\Psi_{ij}V_j\big).
\een
\section{Field perturbations}\lb{s.3}

We consider that the fields are perturbed from a background state where the mass density $\rho$, hydrostatic pressure $p$, specific internal energy $\varepsilon$ and Newtonian gravitational potential $U$ assume constant values, while the background hydrodynamic velocity $V_i$ and the post-Newtonian gravitational potentials $\Phi$, $\Pi_i$, $\Psi_{kk},$ $\Psi_{0i}$ and $\Psi_{00}$ vanish. The background and perturbed  fields are denoted by the sub- and super-scripts 0 and 1, respectively, and the perturbed fields are considered to be small so that only the linear perturbed terms will be considered in the analysis of the present work. The representation of the fields are written as 
\ben\lb{fe1a}
&&\rho(\bx,t)=\rho_0+\rho_1(\bx,t),\qquad V_i(\bx,t)=V_i^1(\bx,t),\\&& U(\bx,t)=U_0+U_1(\bx,t),
\qquad p(\bx,t)=p_0+p_1(\bx,t),\\\lb{fe1b} &&\varepsilon(\bx,t)=\varepsilon_0+\varepsilon_1(\bx,t),\qquad
\Pi_i(\bx,t)=\Pi_i^1(\bx,t),\\\lb{fe1c}
&&\Phi(\bx,t)=\Phi_1(\bx,t),
\qquad \Psi_{0i}(\bx,t)=\Psi_{0i}^1(\bx,t),\\\lb{fe1d} &&\Psi_{00}(\bx,t)=\Psi_{00}^1(\bx,t),\qquad \Psi_{kk}(\bx,t)=\Psi_{kk}^1(\bx,t).
\een

We begin by determining the  linearized expression for the ratio $p/\rho$, yielding
\ben\lb{pph1}
\frac{p}\rho=\frac{p_0}{\rho_0}\left(\frac{1+p_1/p_0}{1+\rho_1/\rho_0}\right)\approx\frac{p_0}{\rho_0}\left(1+\frac{p_1}{p_0}-\frac{\rho_1}{\rho_0}\right).
\een

Next we have to evaluate the perturbed specific internal energy and for that end we make use of  the following  result which comes out from the kinetic theory of relativistic monatomic gases for the specific internal energy $\varepsilon$ (see e.g. \cite{CK}):
\ben\lb{fj5a}
\varepsilon=\frac{3kT}{2m}\left(1+\frac{5kT}{4mc^2}\right)=\frac1{\gamma-1}\frac{p}\rho\left(1+\frac5{6(\gamma-1)}\frac{p}{c^2\rho}\right),
\een
thanks to  the relationship $\varepsilon={p}/{(\gamma-1)\rho}$ where $\gamma=5/3$ is the ratio of the specific heats at constant pressure and constant volume for a monatomic gas.

By taking into account  the expression for the sound speed $c_s^2=dp/d\rho$ the background and perturbed hydrostatic pressure and specific internal energy become
\ben\lb{fe2a}
&&p_0=\frac{c_s^2}\gamma\rho_0,\qquad 
\varepsilon_0=\frac{c_s^2}{\gamma(\gamma-1)}\left(1+\frac5{6\gamma(\gamma-1)}\frac{c_s^2}{c^2}\right),
\\\lb{fe2b}
&&p_1=c_s^2\rho_1,\qquad \varepsilon_1=\frac{c_s^2}{\gamma}\left(1+\frac5{3\gamma(\gamma-1)}\frac{c_s^2}{c^2}\right)\frac{\rho_1}{\rho_0}.
\een

The linearized mass density (\ref{ba2ka}) and its balance equation (\ref{ba2k}) by considering the representations (\ref{fe1a}) -- (\ref{fe1d}) read
\ben\lb{fe3a}
\widetilde \rho=(\rho_0+\rho_1)\left[1+\frac{3U_0}{c^2}+\frac{3U_0^2}{2c^4}\right]+\rho_0\left[\frac{3U_1}{c^2}\left(1+\frac{U_0}{c^2}\right)-\frac{\Psi_{kk}^1}{2c^4}\right],
\\\lb{fe3b}
\bigg[1+\frac{3U_0}{c^2}\bigg(1+\frac{U_0^2}{2c^2}\bigg)\bigg]\bigg[\frac{\partial \rho_1}{\partial t}+\rho_0\frac{\partial V_i^1}{\partial x^i}\bigg]+\frac{3\rho_0}{c^2}\bigg(1+\frac{U_0}{c^2}\bigg)\frac{\partial U_1}{\partial t}-\frac{\rho_0}{2c^4}\frac{\partial \Psi_{kk}^1}{\partial t}=0.
\een
Equation (\ref{fe3b}) can be rewritten as
\ben\lb{fe3c}
\frac{\partial \rho_1}{\partial t}+\rho_0\frac{\partial V_i^1}{\partial x^i}+\frac{3\rho_0}{c^2}\left(1-2\frac{U_0}{c^2}\right)\frac{\partial U_1}{\partial t}-\frac{\rho_0}{2c^4}\frac{\partial \Psi_{kk}^1}{\partial t}=0,
\een
if we multiply it by the first expression within the brackets and keep terms up to the $1/c^4$--order.

In the same way, the linearized momentum  density (\ref{ba4i}) and its balance equation (\ref{ba4h}) by taking into account the representations (\ref{fe1a}) -- (\ref{fe1d}) are
\be
\rho{\mathfrak{V}}_i=\rho_0V_i^1\left\{1+\frac1{c^2}\left(6U_0+\varepsilon_0+\frac{p_0}{\rho_0}\right)+\frac1{c^4}\left[13U_0^2+6U_0\left(\varepsilon_0+\frac{p_0}{\rho_0}\right)\right]\right\}
-\rho_0\Pi_i^1\left[1+\frac1{c^2}\left(4U_0+\varepsilon_0+\frac{p_0}{\rho_0}\right)\right]-\frac{\rho_0}{c^4}\Psi_{0i}^1.
\ee{fe4a}
\ben\no
&&\rho_0\bigg\{1+\frac1{c^2}\left(6U_0+\varepsilon_0+\frac{p_0}{\rho_0}\right)+\frac1{c^4}\bigg[13U^2_0+6U_0\bigg(\varepsilon_0+\frac{p_0}{\rho_0}\bigg)\bigg]\bigg\}\frac{\partial V_i^1}{\partial t}+\bigg(1+\frac{2U_0}{c^2}-\frac{U_0^2}{c^4}\bigg)\frac{\partial p_1}{\partial x^i}
\\\no
&&\qquad-\rho_0\bigg\{1+\frac1{c^2}\bigg(2U_0+\varepsilon_0+\frac{3p_0}{\rho_0}\bigg)
+\frac1{c^4}\bigg[2U_0\bigg(\varepsilon_0+\frac{p_0}{\rho_0}\bigg)-3U_0^2\bigg]\bigg\}\frac{\partial U_1}{\partial x^i}+\frac{\rho_0}{c^4}\bigg(\frac12\frac{\partial \Psi_{00}^1}{\partial x^i}-\frac{\partial \Psi_{0i}^1}{\partial t}\bigg)
\\\lb{fe4b}
&&\qquad-\frac{2\rho_0}{c^2}\bigg[1+\frac1{c^2}\bigg(4U_0+\varepsilon_0+\frac{p_0}{\rho_0}\bigg)\bigg]\bigg(\frac{\partial \Pi_i^1}{\partial t}+2\frac{\partial \Phi_1}{\partial x^i}\bigg)=0.
\een
Following the same methodology as above we multiply (\ref{fe4b}) by the first expression within the brackets and keep terms up to the $1/c^4$--order, yielding
\ben\no
\rho_0\frac{\partial V_i^1}{\partial t}+\bigg\{1-\frac1{c^2}\bigg(4U_0+\varepsilon_0+\frac{p_0}{\rho_0}\bigg)+\frac1{c^4}\bigg[10U_0^2+4U_0\bigg(\varepsilon_0+\frac{p_0}{\rho_0}\bigg)+\bigg(\varepsilon_0+\frac{p_0}{\rho_0}\bigg)^2\bigg]\bigg\}\frac{\partial p_1}{\partial x^i}
\\\lb{fe4c}
-\rho_0\left(1-\frac{4U_0}{c^2}+\frac{8U_0^2}{c^4}\right)\frac{\partial U_1}{\partial x^i}
+\frac{\rho_0}{c^4}\left(\frac12\frac{\partial \Psi_{00}^1}{\partial x^i}-\frac{\partial \Psi_{0i}^1}{\partial t}\right)
-\frac{\rho_0}{c^2}\left(1-\frac{2U_0}{c^2}\right)\left(\frac{\partial \Pi_i^1}{\partial t}+2\frac{\partial \Phi_1}{\partial x^i}\right)=0.
\een

In terms of the perturbed fields the linearized gauge condition (\ref{sp2e}) is expressed as
\ben\lb{fe5}
3\frac{\partial U_1}{\partial t}+\frac{\partial \Pi_i^1}{\partial x^i}+\frac1{c^2}\left(\frac{\partial \Psi_{0i}^1}{\partial x^i}-\frac12\frac{\partial \Psi_{kk}^1}{\partial t}\right)=0.
\een

For the Poisson equations   we make use of "Jeans swindle" and assume that (\ref{sp2a}) -- (\ref{sp2d})  are only valid for the perturbed fields, so that the linearized Poisson equations read
\ben\lb{fe6a}
&&\nabla^2U_1=-4\pi G\rho_1,
\\\lb{fe6b}
&&\nabla^2\Phi_1=-4\pi G\rho_0\left[U_1+\frac{\varepsilon_1}2+\frac{3p_1}{2\rho_0}+\frac{2U_0+\varepsilon}{2\rho_0}\rho_1\right],
\\\lb{fe6c}
&&\nabla^2\Pi_i^1=-16\pi G\rho_0 V_i^1+\frac{\partial^2U_1}{\partial t\partial x^i},
\\\lb{fe6d}
&&\nabla \Psi_{kk}^1=32\pi G\rho_0\left[4U_1+\varepsilon_1+\frac{4U_0+\varepsilon_0}{\rho_0}\rho_1\right],
\\\lb{fe6e}
&&\nabla^2\Psi_{0i}^1=-16\pi G\rho_0\left[V_i^1\left(\varepsilon_0+\frac{p_0}{\rho_0}+4U_0\right)-\frac{\Pi_i^1}2\right],
\\\lb{fe6f}
&&\nabla^2\Psi_{00}^1=-32\pi G\rho_0\left(U_0U_1+\Phi_1+\frac{U_0^2}{2\rho_0}\rho_1\right).
\een

The time derivative of the perturbed mass density balance equation (\ref{fe3c}) and the spatial divergence of the perturbed momentum density balance equation (\ref{fe4c}) lead to
\ben\lb{fe7a}
\frac{\partial^2\rho_1}{\partial t^2}+\rho_0\frac{\partial^2 V_i^1}{\partial t \partial x^i}+\frac{3\rho_0}{c^2}\left(1-\frac{2U_0}{c^2}\right)\frac{\partial^2U_1}{\partial t^2}-\frac{\rho_0}{2c^4}\frac{\partial^2\Psi_{kk}^1}{\partial t^2}=0,
\\\no
\rho_0\frac{\partial^2 V_i^1}{\partial t\partial x^i}+\bigg\{1-\frac1{c^2}\bigg(4U_0+\varepsilon_0+\frac{p_0}{\rho_0}\bigg)+\frac1{c^4}\bigg[10U_0^2+4U_0\bigg(\varepsilon_0+\frac{p_0}{\rho_0}\bigg)+\bigg(\varepsilon_0+\frac{p_0}{\rho_0}\bigg)^2\bigg]\bigg\}c_s^2\nabla^2\rho_1
\\\lb{fe7b}
-\rho_0\bigg(1-\frac{4U_0}{c^2}+\frac{8U_0^2}{c^4}\bigg)\nabla^2 U_1
+\frac{\rho_0}{c^4}\bigg(\frac{\nabla^2\Psi_{00}^1}2-\frac{\partial^2 \Psi_{0i}^1}{\partial t\partial x^i}\bigg)
-\frac{\rho_0}{c^2}\bigg(1-\frac{2U_0}{c^2}\bigg)\bigg(\frac{\partial^2 \Pi_i^1}{\partial t\partial x^i}+2\nabla^2 \Phi_1\bigg)=0.
\een

Now by eliminating the  perturbed hydrodynamic velocity $V_i^1$ from (\ref{fe7a}) by using (\ref{fe7b}) we get 
\ben\no
\frac{\partial^2\rho_1}{\partial t^2}-\bigg\{1-\frac1{c^2}\bigg(4U_0+\varepsilon_0+\frac{p_0}{\rho_0}\bigg)+\frac1{c^4}\bigg[10U_0^2+4U_0\bigg(\varepsilon_0+\frac{p_0}{\rho_0}\bigg)+\bigg(\varepsilon_0+\frac{p_0}{\rho_0}\bigg)^2\bigg]\bigg\}c_s^2\nabla^2\rho_1
\\\no
+\rho_0\bigg(1-\frac{4U_0}{c^2}+\frac{8U_0^2}{c^4}\bigg)\nabla^2 U_1+\frac{2\rho_0}{c^2}\bigg(1-\frac{2U_0}{c^2}\bigg)\nabla^2 \Phi_1-\frac{\rho_0}{2c^4}\nabla^2\Psi_{00}^1
\\\lb{fe8}
+\frac{\rho_0}{c^2}\frac{\partial}{\partial t}\underline{\bigg[3\frac{\partial U_1}{\partial t}+\frac{\partial \Pi_i^1}{\partial x^i}+\frac1{c^2}\bigg(\frac{\partial\Psi_{0i}^1}{\partial x^i}-\frac12\frac{\partial \Psi_{kk}^1}{\partial t}\bigg)}\bigg]-\frac{2\rho_0U_0}{c^4}\frac{\partial}{\partial t}\underline{\bigg[3\frac{\partial U_1}{\partial t}+\frac{\partial \Pi_i^1}{\partial x^i}\bigg]}=0.
\een
Note that the underlined terms vanish thanks to the gauge condition (\ref{fe5}) in the $\mathcal{O}(c^{-2})$ and $\mathcal{O}(c^{-4})$ orders.

Equations (\ref{fe6a}),   (\ref{fe6b}), (\ref{fe6f}) and (\ref{fe8}) represent a system of differential equations for the determination of the perturbations $U_1$, $\Phi_1$, $\Psi_{00}^1$ and $\rho_1$. This system of differential equations will be analysed in the next section by considering a plane wave representation for the perturbed fields.

\section{Plane wave representations}\lb{s.4}

We consider that the perturbed fields  $U_1$, $\Phi_1$, $\Psi_{00}^1$ and $\rho_1$ are represented by plane waves of wave number vector $\bk$ and angular frequency $\omega$, namely
\ben\lb{fj7d}
\rho_1(\bx,t)=\overline{\rho}e^{\left[i\left(\bk\cdot \bx-\omega t\right)\right]},\qquad U_1(\bx,t)=\overline{U}e^{\left[i\left(\bk\cdot \bx-\omega t\right)\right]},
\\\lb{fj7e}
\Phi_1(\bx,t)=\overline{\Phi}e^{\left[i\left(\bk\cdot \bx-\omega t\right)\right]},\quad \Psi_{00}^1(\bx,t)=\overline\Psi_{00}e^{\left[i\left(\bk\cdot \bx-\omega t\right)\right]},
\een
where $\overline\rho$, $\overline U$, $\overline\Phi$ and $\overline\Psi_{00}$ are small amplitudes.

Insertion of the plane wave representations  (\ref{fj7d}) and (\ref{fj7e}) into  the equations  (\ref{fe6a}),   (\ref{fe6b}), (\ref{fe6f}) and (\ref{fe8}), yield the following system of algebraic equations for the amplitudes $\overline\rho$, $\overline U$, $\overline\Phi$ and $\overline\Psi_{00}$:
\ben\lb{fj7b}
&&\kappa^2\overline U=4\pi G\overline\rho,\qquad \kappa^2\overline\Psi_{00}=16\pi G\rho_0\left[2\left(\overline\Phi+U_0\overline U\right)+U_0^2\overline\rho\right],
\\\lb{fj7c}
&&\kappa^2\overline\Phi=4\pi G\rho_0\overline U+2\pi G\left\{2U_0+\frac{c_s^2}{\gamma-1}\left[3\gamma-2+\frac{5(2\gamma-1)}{6\gamma^2(\gamma-1)
}\frac{c_s^2}{c^2}\right]\right\}\overline\rho,
\\\no
&&\bigg\{\omega^2-c_s^2\kappa^2\bigg[1-\frac1{c^2}\bigg(4U_0+\varepsilon_0+\frac{p_0}{\rho_0}\bigg)+\frac1{c^4}\bigg(10U_0^2+4U_0\bigg(\varepsilon_0+\frac{p_0}{\rho_0}\bigg)+\bigg(\varepsilon_0+\frac{p_0}{\rho_0}\bigg)^2\bigg)\bigg]\bigg\}\overline\rho
\\\lb{fj7a}
&&\qquad+\rho_0\kappa^2\bigg[\bigg(1-\frac{4U_0}{c^2}+\frac{8U_0^2}{c^4}\bigg)\overline U
+\frac{2}{c^2}\bigg(1-\frac{2U_0}{c^2}\bigg)\overline\Phi-\frac{\overline\Psi_{00}}{2c^4}\bigg]=0.
\een
In the Poisson equations (\ref{fj7b}) and (\ref{fj7c}) we have used the relations  (\ref{fe2a}) and (\ref{fe2b}) and introduced the modulus of wave number vector $\kappa=\sqrt{\bk\cdot\bk}$.

The elimination  from (\ref{fj7a}) of the amplitudes $\overline U$, $\overline \Phi$ and $\overline \Psi_{00}$  by using the Poisson equations (\ref{fj7b}) and  (\ref{fj7c}) together with the relations (\ref{fe2a})  leads to the following dispersion relation 
\ben\no
&&\omega_*^2=\kappa_*^2-1-\frac{c_s^2}{c^2}\bigg[\bigg(\frac{4U_0}{c_s^2}+\frac{1}{\gamma-1}\bigg)\kappa_*^2+\frac2{\kappa_*^2}-\frac{2U_0}{c_s^2}+\frac{3\gamma-2}{\gamma-1}\bigg]-\frac{c_s^4}{c^4}\bigg[\bigg(\frac{(5-6\gamma)}{6\gamma^2(\gamma-1)^2}-\frac{4U_0}{c_s^2(\gamma-1)}-\frac{10U_0^2}{c_s^4}\bigg)\kappa_*^2
\\\lb{fj8a}
&&\qquad-\frac{12U_0}{c_s^2\kappa_*^2}
-\frac{2(3\gamma-2)}{(\gamma-1)\kappa_*^2}\bigg(1+\frac{U_0}{c_s^2}\bigg)+\frac{2U_0^2}{c_s^4}+\frac{5(2\gamma-1)}{6(\gamma-1)^2\gamma^2}-\frac4{\kappa_*^4}\bigg],
\een
where only the terms up to the $1/c^4$ order were considered. Furthermore, in (\ref{fj8a})  we have introduce the dimensionless angular frequency $\omega_*$ and the dimensionless wavenumber $\kappa_*$ defined by
\ben
\omega_*=\frac{\omega}{\sqrt{4\pi G \rho_0}},\qquad \kappa_*=\frac\kappa{\kappa_J}, \qquad \hbox{where}\qquad \kappa_J=\frac{\sqrt{4\pi G \rho_0}}{c_s},
\een
is the Jeans wave number.

From the solution of  (\ref{fj8a}) for  the dimensionless angular frequency $\omega_*$ one has two distinct behaviors: if  $\kappa_*>1$, $\omega_*$ assumes real values and the perturbations will propagate as  harmonic waves in time, while if $\kappa_*<1$,  $\omega_*$  acquires pure imaginary values and  the perturbations will grow or decay in time. The one which grows with time is associated with the Jeans instability.
Hence, by considering $\omega_*=0$  in  (\ref{fj8a}) we can determine the value of  $\kappa_*$ where $\omega_*$ changes from the real value to the pure imaginary value. The real solution of  (\ref{fj8a}) for $\kappa_*$ when $\omega_*=0$ by considering only terms up to the $1/c^4$ order is
\ben\no
&&\kappa_*=1+\bigg[\frac{5\gamma-3}{2(\gamma-1)}+\frac{U_0}{c_s^2}\bigg]\frac{c_s^2}{c^2}+\bigg[\frac{20-9\gamma(\gamma-1)(35\gamma-27)}{24\gamma(\gamma-1)^2}+\frac{9-7\gamma}{2(\gamma-1)}\frac{U_0}{c_s^2}-\frac{U_0^2}{2c_s^4}\bigg]\frac{c_s^4}{c^4}
\\\lb{fj8b}
&&\qquad=1+\left[4+\frac{U_0}{c_s^2}\right]\frac{c_s^2}{c^2}-\underline{\left[\frac{33}2+2\frac{U_0}{c_s^2}+\frac{U_0^2}{2c_s^4}\right]\frac{c_s^4}{c^4}}.\quad
\een
In the last equality we have introduced the value $\gamma=5/3$. In the above equation the term in $c_s^2/c^2$  represents the  contribution of the first  post-Newtonian approximation to the dimensionless modulus of the wave number and was obtained in \cite{NH,gg1}. The underlined term in $c_s^4/c^4$ corresponds to the second post-Newtonian contribution and it is interesting to note that this contribution is negative. 

 Let us analyse the  Jeans  mass which corresponds to the minimum amount  of  mass  for an overdensity  to  initiate  the  gravitational collapse.  The Jeans mass represents the mass contained in a sphere of radius equal to the wavelength of the perturbation. Hence, if $M_J^{PN}$ denotes the mass corresponding to the post-Newtonian wavelength  and $M_J^N$  the Newtonian one, we can build the ratio by taking into account (\ref{fj8b}), yielding 
\be
\frac{M_J^{PN}}{M_J^N}=\frac{\lambda^3}{\lambda_J^3}=\frac{\kappa_J^3}{\kappa^3}=\frac1{\kappa_*}\approx 1-3\frac{c_s^2}{c^2}\left[4+\frac{U_0}{c_s^2}\right]+\frac12\frac{c_s^4}{c^4}\left[291+108\frac{U_0}{c_s^2}+15\frac{U_0^2}{c_s^4}\right].
\ee{jpn20}

We note  from  (\ref{jpn20})  that the mass for an overdensity needed to begin the gravitational collapse  has contributions from the first and second  post-Newtonian approximations and that the presence of the background Newtonian potential $U_0$ has influence on it. If we consider  the virial theorem and approximate  the square of the sound speed  with the  background Newtonian gravitational potential $U_0\approx c_s^2$, we get that  (\ref{jpn20}) can be written as 
\ben\lb{jpn20d}
\frac{M_J^{PN}-M_J^N}{M_J^N}= -15\frac{c_s^2}{c^2}+207\frac{c_s^4}{c^4},
\een
which shows  the relative difference  of the post-Newtonian and Newtonian  Jeans masses. While the relative difference of the first post-Newtonian approximation is negative, the one of the second post-Newtonian approximation is positive. The correction of the first post-Newtonian approximation is the preponderant one, indeed if we consider that $c_s=5\%\,c$ the ratio ${M_J^{PN}}/{M_J^N}= 0.9625$ by considering only the first post-Newtonian approximation and ${M_J^{PN}}/{M_J^N}= 0.9638$ by considering the first and second post-Newtonian approximations.  Hence, we may conclude that the two contributions  imply a smaller mass  needed for an overdensity to initiate  the gravitational collapse with respect to the one given by the Newtonian theory.

\section{Conclusions}\lb{s.6}
To sum up: in this work the equations of the second post-Newtonian approximation to Einstein's field equations were analysed. The starting point was the  hydrodynamic equations for the mass and momentum densities which were coupled with six Poisson equations for the Newtonian and post-Newtonian gravitational potentials. The basic fields and gravitational potentials were perturbed from a background state and plane wave representations for the perturbations imply a system of algebraic equations whose solution is a dispersion relation  where the condition related with the Jean instability emerges.  The influence of the first and second post-Newtonian approximations in the Jeans mass -- which is related with the minimum mass necessary for an overdensity to initiate the gravitational collapse -- are determined. It was shown that the relative difference  of the first post-Newtonian and Newtonian  Jeans masses is negative, while the one of the second post-Newtonian is positive. The contributions of the first and second approximations lead to a smaller Jeans mass in comparison to the one given by the Newtonian theory.

\section*{Acknowledgments} {This work was supported by Conselho Nacional de Desenvolvimento Cient\'{i}fico e Tecnol\'{o}gico (CNPq), grant No.  304054/2019-4.}

\end{document}